\documentclass[twocolumn,showpacs,preprintnumbers,amsmath,amssymb,floatfix]{revtex4}
\usepackage{graphicx}
\usepackage{dcolumn}
\usepackage[latin1]{inputenc}

\begin{document}

\preprint{Preprint}

\title {An alternate theoretical approach to diffusion bonding}

\author{Miguel Lagos}
\email{mlagos@utalca.cl}
\affiliation{Facultad de Ingenier\'\i a, Universidad de Talca,
Campus Los Niches, Curic\'o, Chile}

\author{C\'esar Retamal}
\email{ceretamal@utalca.cl}
\affiliation{Facultad de Ingenier\'\i a, Universidad de Talca,
Campus Los Niches, Curic\'o, Chile}
\affiliation{Universit\'e de Paris Ouest Nanterre La D\'efense,
Laboratoire de Thermique Interfaces Environnement (LTIE), EA4415, GTE,
50 rue de Sèvres, 92410 Ville d'Avray, France}

\date{June 15, 2010}

\begin{abstract}
On the basis of a previous theoretical approach to the plastic flow of
highly refined materials, a physical explanation for diffusion bonding
is essayed, which yields closed--form equations relating the bonding
progress with time, temperature, applied pressure and the constants
characterizing the material. Excellent agreement with experiment is
attained, with no adjustable parameter. In the novel scheme, diffusion 
bonding is caused by the interpenetration of the two sufaces at the
grain level. The process is driven by the strong tensile stress field
induced in the plane of the interface by the plastic deformation in
the normal direction. The grain boundaries of each joining surface
yield to host grains of the other surface, releasing this way the
internally generated tensile stresses. Voids gradually close with the
increment of the interpenetrated areas. In this scheme bonding is not
a matter of contacting and atomic interdiffusion, but of grain
exchange.

\end{abstract}


\maketitle

Diffusion bonding (DB) is a solid state high temperature process for
making a monolithic joint through matter transportation across the
interface between the materials being joined, with no melting at the
bond line. It is a costly time--consuming technique for producing top
quality bonds, so that neither metallurgical discontinuities nor voids
can be detected all over the former interface. Though not worthy for
mass production assembly lines yet, millions people are benefited
every day by diffusion bonded artifacts. The titanium hollow blades,
their internal reinforcing honeycomb structure, and the central disc
of the wide chord fan of a present day turbofan commercial aeroengine,
which generates about 80\% of the total engine thrust at take--off,
integrate a single component diffusion bonded unit. On the basis of a
recently published theoretical approach to the plastic flow of highly
refined materials, we essay here a physical explanation for DB and
derive a closed--form equation relating bonding time, temperature,
applied pressure and the constants characterising the material, which
exhibits excellent agreement with available experimental data, and
with no adjustable parameter. In our theoretical approach, DB has
little connection, if any, with atomic diffusion.

For two decades, DB combined with superplastic forming (SPF) has been
a standard technique for producing aircraft structural components and
engine parts, particularly because SPF yields practically no residual
stress and neither spring back from die after releasing the forming
pressure. Besides, DB eliminates fasteners, yielding stronger more
reliable unions, saving weigth and allowing better design
\cite{Hefti,Nieh}.  However, the suitability of the process for mass
production is yet questionable, mainly by the long bonding times
involved and because DB does work just for a short list of known
materials. The precise knowledge of the microscopic mechanisms taking
part in the process, leading to the physical equations governing its
evolution, would be of great value as scientific knowledge, as well as
a tool for the search of new improved materials and process
optimisation.

The theoretical literature on the subject assumes that any pair of
clean metallic surfaces will bond if they are brought together within
the range of interatomic forces. Bonding happens by atomic
interdiffusion across the interfaces at the contacting surface
sectors, which initially are isolated and scarce by the surface
roughness. Contacting area gradually increases by plastic deformation
and atomic interdiffusion. The large differences in bonding ability
observed for the various alloys are ascribed to the chemical and
mechanical stability of their surface oxide layers, which may obstruct
atomic exchange \cite{Hefti,Pilling1,Ridley}.

In our scheme the process is quite different. It takes place mainly by
the active migration of grains that cross the joint line in the two
senses, rather than intergrain diffusion of individual atoms. Grain
exchange is much more efficient and demands surface proximity of the
order of just the grain size, instead of the much smaller atomic
distances. The driving force for the interfacial grain exchange arises
from the internal stress which plastic deformation necessarily induces
in the plane normal to the flow direction.

DB has been observed in highly refined materials constituted by small
equiaxed grains, whose plastic deformation conserves the grain mean
size $d$ and overall shapes. Thus, grains slide past each other over
relatively long distances, accommodating their shapes to preserve
matter continuity, with grain boundaries maintaining mechanical
integrity and coherence.  Hence grains evolve by the action of two
kinds of forces that may be of very different strength: those making
their boundaries to slide and the induced local forces responsible for
the continuous grain reshaping. The stronger one will determine the
plastic flow rate.

Recent papers show that excellent agreement with experiment is
attained from ascribing the rate control to the intergrain forces that
make adjacent grains to slide
\cite{Lagos1,Lagos2,LagosRetamal,LagosRetamal1}. These forces have
been examined by theory \cite{Lagos1,Lagos2}, Montecarlo computer
simulation
\cite{Bellon}, experiment \cite{Fukutomi} and molecular dynamics
simulation \cite{Qi}. Said studies clearly show that the onset of
grain boundary sliding is associated to a boundary mechanical
instability which demands that the shear stress in the plane of the
common boundary of the sliding grains be higher than a critical value
$\tau_c$. Then the grains start sliding with a relative velocity
$\Delta\vec v$ which may be proportional to: (a) The difference
between the in--plane overcritical shear stress and the threshold
stress $\tau_c$.  (b) The bare in--plane shear stress, and $\Delta\vec
v$ jumps from zero to the proper value when $\tau_c$ is surpassed. We
refer to these two options as force models A and B, respectively
\cite{LagosRetamal1}.

As the two models can be fully worked out without additional
assumptions, comparison with experiment determines what model is the
right one for a specific system. It has been shown that model A works
for a series of aluminium and titanium alloys
\cite{Lagos2,LagosRetamal}, fitting the experimental data within the
experimental uncertainties, and model B gives very good results for
the steel Avesta 2304 \cite{LagosRetamal1}.

The technical details for deriving the plastic flow of the
polycrystalline medium from the hypotheses stated above are in the
literature \cite{Lagos1,Lagos2,LagosRetamal,LagosRetamal1}.
Summarising, given the stress tensor $(\sigma_{ij})$, $i,j=x,y,z$, the
force law A or B determines the relative velocities $\Delta\vec
v(\theta ,\phi)$ between adjacent grains. Angles $\theta$ and $\phi$
characterise the grain boundary orientation. One can readily realise
that

\begin{equation}
\frac{\partial v_i}{\partial x_j}
=\frac{1}{d}\langle\Delta v_i\rangle_j \, ,
\label{E1}
\end{equation}

\noindent
where $\vec v$ is the velocity field of the material medium and symbol
$\langle\dots\rangle_j$ stands for the average over all boundary
orientations along the $x_j$ axis in the positive sense. (Notice that
we denote either $i,j=x,y,z$ or $x_i=x,y,z$, $i=1,2,3$). Making the
averages in explicit way, closed--form equations relating $\partial
v_i/\partial x_j$ with the stresses are derived from Eq.~(\ref{E1}).
Rotation invariants like $\nabla\cdot\vec v$ and $\nabla\times\vec v$,
or the components $\dot\varepsilon_{ij}$ of the strain rate tensor,
which is the symmetric part of tensor (\ref{E1}), can be constructed
with them.

For uniaxial external stress $\sigma$ along the $z$ axis and
cylindrical symmetry with respect to the same axis, so that
$\sigma_{xx} =\sigma_{yy}\equiv\sigma_\perp$ and $\sigma_{ij}=0$ for
$i\ne j$, the strain rate along $z$ is shown to be

\begin{equation}
\dot\varepsilon =\frac{\partial v_z}{\partial z}=
s\,\frac{\tau_c\mathcal{Q}(p)}{2d}\left[ \cot (2\theta_c)
+\alpha\left( 2\theta_c-\frac{\pi}{2}\right)\right] ,
\label{E2}
\end{equation}

\noindent
and

\begin{equation}
\begin{aligned}
\nabla\cdot\vec v =
&-s\frac{\tau_c\mathcal{Q}(p)}{2d}
\bigg[ \frac{1 -\cos(2\theta_c)}{\sin (2\theta_c)}
-2\theta_c\bigg(\alpha +\frac{2}{\pi \sin (2\theta_c)}\bigg)\\
&+\frac{2}{\pi}(1-2\alpha)\cos (2\theta_c)+\alpha\frac{\pi}{2}\bigg]
\, ,
\label{E3}
\end{aligned}
\end{equation}

\noindent
where $s$ is the sign of the deviatoric stress $\sigma +p$, being $p$
the pressure $p=-(\sigma +2\sigma_\perp)/3$, and the coefficient
$\mathcal{Q}(p)$ is the proportionality factor between the relative
speed of adjacent grains and the resolved overcritical shear
stress. Temperature $T$ apart, $\mathcal{Q}(p)$ depends only on the
pressure invariant $p$ because it must keep unchanged for any boundary
orientation. The auxiliary variable $\theta_c$ is defined by

\begin{equation}
\sin(2\theta_c)=\frac{4\tau_c}{3|\sigma +p|}
\label{E4}
\end{equation}

\noindent
and $\alpha$ takes the values 1 or 0 for force models A or B, respectively.

Eq.~(\ref{E3}) discloses a remarkable consequence of the
existence of a finite threshold stress for grain sliding. As
$\nabla\cdot\vec v =\dot V/V\ne 0\,$, where $\dot V/V$ is the dilation
rate per unit volume, the specific volume is not conserved in plastic
flow if $\tau_c\ne 0$. As grain volume variations can only be elastic,
one can recall Hooke's law and write $\nabla\cdot\vec v=-(1/B)\dot p$,
where $B$ is the bulk elastic modulus.  Combining this with equation
(\ref{E3}) it follows the equation

\begin{equation}
\begin{aligned}
\dot p=
&s\frac{B\tau_c\mathcal{Q}(p)}{2d}
\bigg[ \frac{1 -\cos(2\theta_c)}{\sin (2\theta_c)}
-2\theta_c\bigg(\alpha +\frac{2}{\pi \sin (2\theta_c)}\bigg)\\
&+\frac{2}{\pi}(1-2\alpha)\cos (2\theta_c)+\alpha\frac{\pi}{2}\bigg]
\label{E5}
\end{aligned}
\end{equation}

\noindent
for the evolution of $p$ on plastic flow.

Pressure $p$ monotonically increases or decreases for plastic axial
stretching ($s>0$) or shrinking ($s<0$), respectively. As $\sigma$ is
the externally applied stress, the increase of $|p|$ must be
attributed to the variation of

\begin{equation}
\sigma_\perp =-\frac{1}{2}(\sigma +3p),
\label{E6}
\end{equation}

\noindent
which represent an internal radial stress field normal to the plastic
flow direction. Settling $\sigma_\perp =0$ (or $p=-\sigma/3$) at
start, when the strain $\varepsilon =0$, is a natural initial
condition. However, as the deformation proceeds, $\sigma_\perp$
becomes finite, taking negative values on stretching, and hence
contributing to neck formation, and positive ones on axial
compression. Both $p$ and $\sigma_\perp$ grow at a high rate because
$B\sim 10^5\,\text{MPa}$. If the plastic flow is interrupted, the
deformation induced stress field $\sigma_\perp$ will remain as a
residual stress.

We must remark that the equations written above assume cylindrical
symmetry. In strict rigor, they hold for an axially symmetric sample
or at any point of an infinite medium.  Symmetry breaking surfaces
make the problem much more complex. However, there is a practical
situation for which Eqs.~(\ref{E2}) and (\ref{E5}) hold strictly:
if the material is finite in the $z$ direction, but infinite in the
$xy$ plane, one can anyway assume cylindrical symmetry at any point of
it. Hence, Eqs.~(\ref{E2}) and (\ref{E5}) are obeyed by an
extended flat plate subjected to a compressive stress $\sigma$ normal
to its plane, and strong enough to produce plastic deformation. In
such case the material has no means to relax the tensile stress
$\sigma_\perp$, which progressively builds up in the $xy$ plane upon
plastic compression in the $z$ direction.

However, there is again a practical situation for which the extended
plate is able to relax the induced tensile stress, at least partially,
at the surfaces. If we compress together two plates with surface
irregularities of the order of the grain size $d$, the grains of one
of them may cede to $\sigma_\perp$ and open up if the consequent
interstice is simultaneously filled by one or more grains provided by
the other surface.  The process is reciprocal, and the
interpenetration of the two surfaces at the grain level provides a
mechanism to release the tensile cumulative stress $\sigma_\perp$.
The interpenetration of the contacting surfaces makes their areas to
increase, filling this way the voids left by the surface
irregularities.

The rate of bonding is dictated by the growing rate of the total
interpenetrating surface area $A$.  A good criterion is to assume that
$A$ varies at a speed that keeps $\sigma_\perp =0$. We have that $\dot
A/A=\nabla\cdot\vec v-\partial v_z/\partial z$. Hence $\dot A/A$ is
obtained substracting Eq.~(\ref{E2}) from Eq.~(\ref{E3}),
with $s=-1$ and $p=-\sigma /3$, or, equivalently, $\sin (2\theta_c)=
2\tau_c/|\sigma|$. Additionally, the coefficient $\mathcal{Q}(p)$ was
studied in Refs.~\onlinecite{Lagos1} and \onlinecite{Lagos2}. The
mechanisms for grain boundary sliding considered there give

\begin{equation}
\frac{\mathcal{Q}(p)}{4d}=C_0\frac{\Omega^*}{k_BT}
\exp\left( -\frac{\epsilon_0+\Omega^*p}{k_BT}\right)
\label{E7},
\end{equation}
 
\noindent
where $k_B$ is the Boltzmann constant, the coefficient $C_0$ depends
only on the grain size $d$, the constant $\epsilon_0$ is the energy
necessary for evaporating a crystal vacancy from the grain boundary,
and $\Omega^*$ measures the sensitivity of this energy to
stress. Combining all this, one has that
   
\begin{equation}
\begin{aligned}
\lambda =&\bigg(\frac{\dot A}A\bigg)_0=
2C_0\frac{\Omega^*\tau_c}{k_BT}\bigg[ \frac{|\sigma |}{2\tau_c}
-\frac{2}{\pi}\frac{|\sigma |}{2\tau_c}
\arcsin\bigg(\frac{2\tau_c}{|\sigma|}\bigg)\\
&+\frac{2}{\pi}(1-2\alpha)
\sqrt{1-\bigg(\frac{2\tau_c}{|\sigma |}\bigg)^2}\,\bigg]\\
&\times\exp\bigg(-\frac{\varepsilon_0+\Omega^*|\sigma|/3}{k_BT}\bigg)
\qquad (\sigma_\perp =0),
\label{E8}
\end{aligned}
\end{equation}

\noindent
where the subscript of $(\dot A/A)_0$ indicates that equation
(\ref{E8}) gives the initial grow rate of the bonded area fraction,
when $\sigma_\perp =0$.

The condition $\sigma_\perp =0$ will cease only once voids be almost
filled.  Hence, what matters in practice is the initial binding rate
$\lambda$, and it can be assumed that $\dot A$ vanishes when
$\sigma_\perp$ becomes finite because $A$ has no free room to grow.
On this basis one can write the equation

\begin{equation}
\dot A=\lambda\frac{A_1}{A_0}\left( A_1-A\right),
\label{E9}
\end{equation}
 
\noindent
where $A_1$ is the total area to be bound, $A_0$ is the initial
contact area, constituted by $A_1/A_0$ small islands which are assumed
already joined. $A$ is the bonded area at time $t$ and $A_1-A$ the
unbound area. The coefficient $\lambda$ measures the intrinsic ability
of the material for making DB joints. Eq.~(\ref{E9}) has the
solution

\begin{equation}
\frac{A}{A_1}=1-\exp\left( -\lambda\frac{A_1}{A_0}t\right),
\label{E10}
\end{equation}
 
\noindent
which gives the time dependent area fraction bonded.

\begin{figure}[h!]
\includegraphics[width=8.5cm]{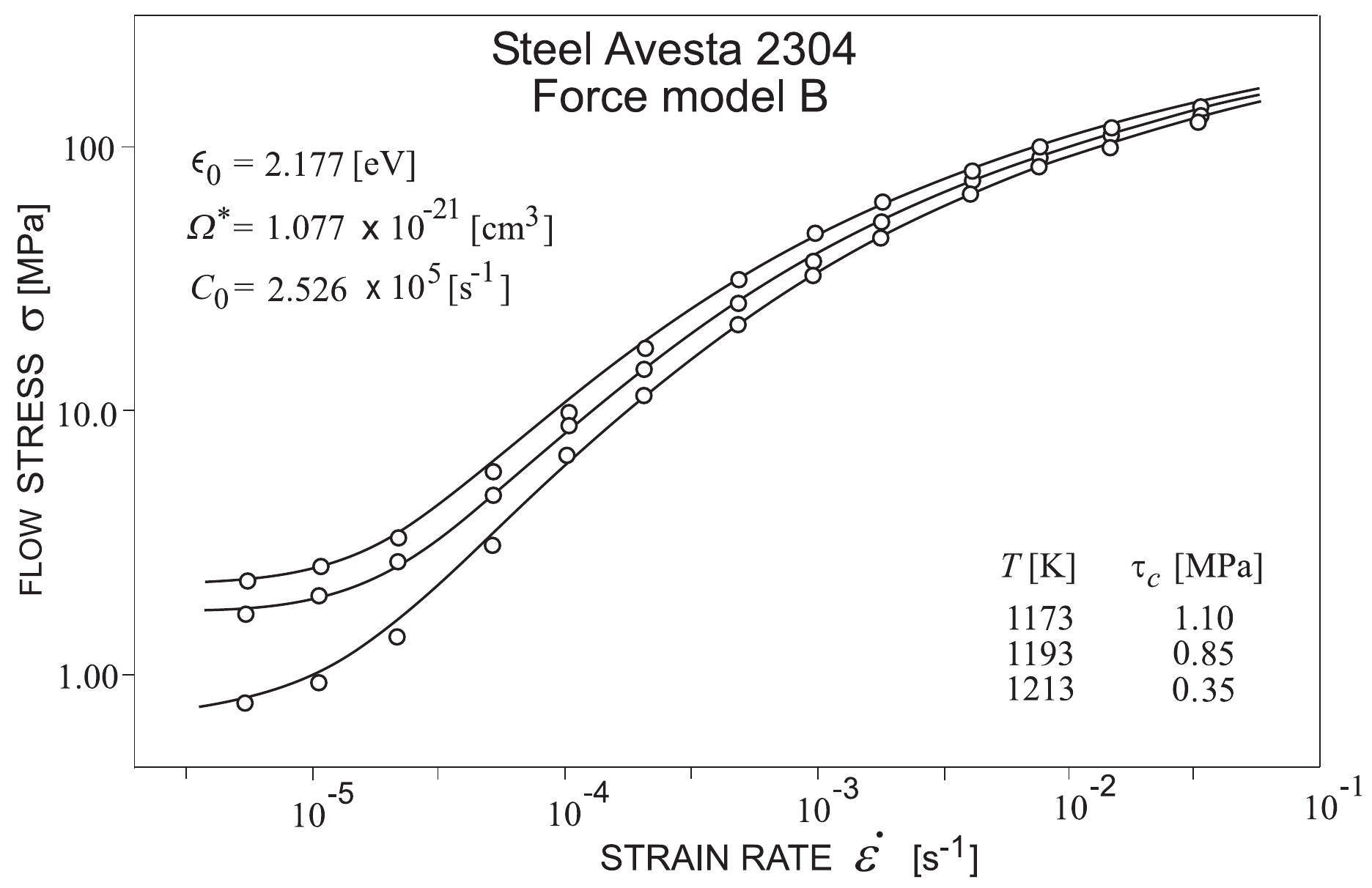}
\caption{\label{Fig1} Flow stress {\it versus} strain rate for the
steel Avesta 2304 at $T=1173$, $1193$ and $1213\,\text{K}$. Circles
represent experimental data taken from Ref.~\onlinecite{Pilling2} and
solid lines depict Eq.~(\ref{E2}) with $\alpha =0$ and the
constants shown in the insets. The parameter $\tau_c$ exhibits some
temperature dependence.}
\end{figure}

\begin{figure}[h!]
\includegraphics[width=8.5cm]{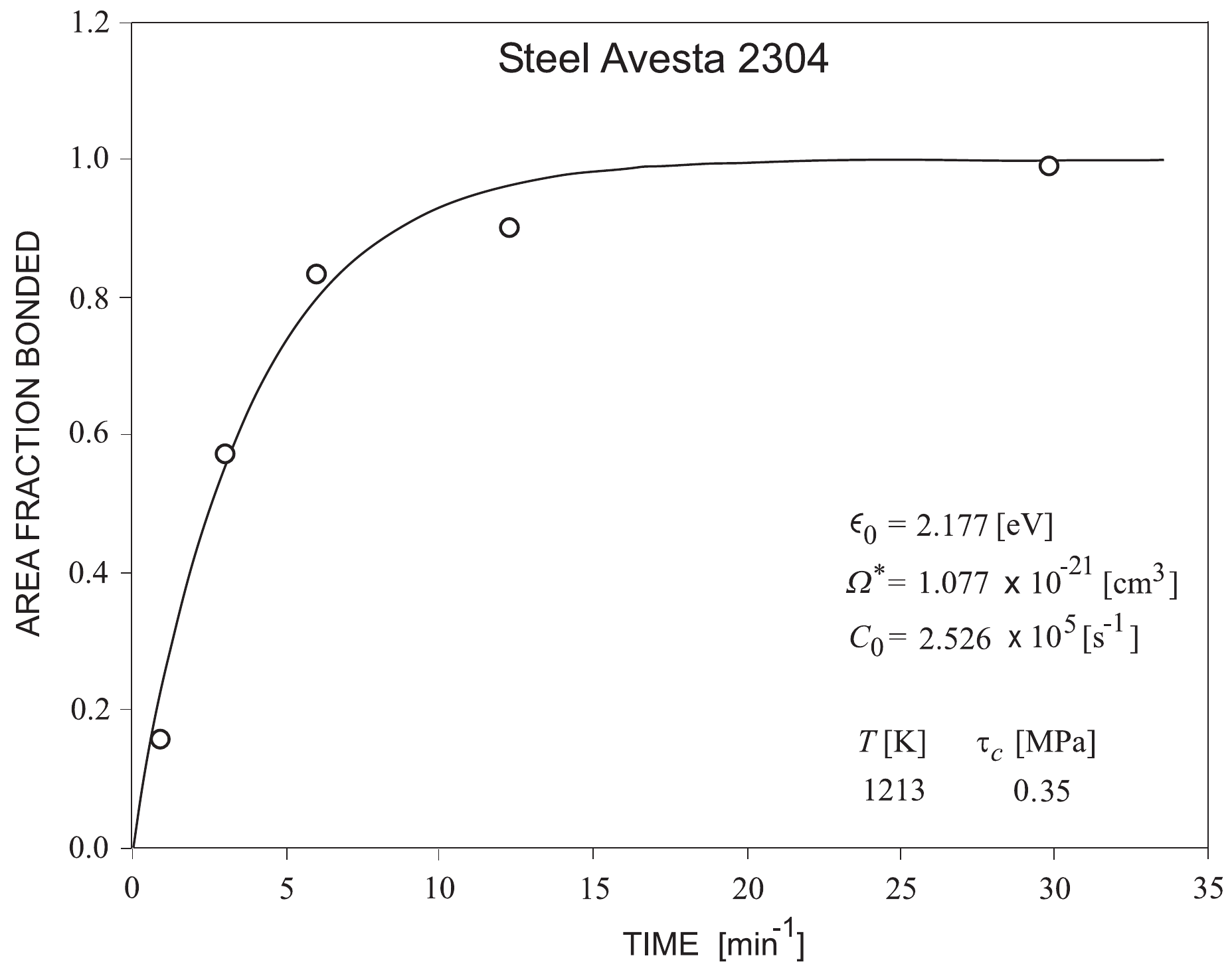}
\caption{\label{Fig2} The variation of the area fraction bonded with
time for steel Avesta 2304 at $T=1213\,\text{K}$ and $\sigma
=3.0\,\text{MPa}$.  Circles represent experimental data taken from
Ref.~\onlinecite{Pilling2}.  The solid line represents equation
(\ref{E10}) with $\lambda$ given by Eq.~(\ref{E8}) and the
material constants obtained from the fit shown in Fig.~\ref{Fig1}. The
theoretical curve has no adjustable parameter.}
\end{figure}

Pilling, Wang and Ridley \cite{Pilling2} authored one of the very
scarce experimental observations of the time progression of DB
available in the literature. Together with the time evolution
$A(t)/A_1$ of two bonding surfaces of the steel Avesta 2304 at
$T=1213\,\text{K}$ and $\sigma =3.0\,\text{MPa}$, they reported also
the flow stress $\sigma$ for plastic stretching versus strain rate
$\dot\varepsilon$ at several temperatures. Therefore, their data
allows us to test Eq.~(\ref{E2}), and Eqs.~(\ref{E8}) and
(\ref{E10}) independently. Fig.~\ref{Fig1} shows the fit given by
Eq.~(\ref{E2}) to the flow stress data ($\varepsilon\approx 0$)
at three temperatures, which gives us precise values for the constants
$\epsilon_0$, $C_0$, $\Omega^*$ and $\tau_c$ of the material.

Inserting these material constants in Eq.~(\ref{E8}), together
with $\sigma =3.0\,\text{MPa}$ and $T=1213\,\text{K}$, it is obtained
that $\lambda =4.11\times 10^{-5}\,\text{s}^{-1}=2.47\times
10^{-3}\,\text{min}^{-1}$.  On the other hand, Pilling {\it et al.}
\cite{Pilling2} reported that the two joining steel surfaces in the DB
experiment initially display very small contact sectors, of less than
$4\,\mu\text{m}$, distant $l=50\, -\, 55\,\mu\text{m}$ between
them. We take for the ratio $A_1/A_0=(l/d)^2$ because the size of the
initial bound spots cannot be smaller than the grain size
$d=5\,\mu\text{m}$. Therefore, taking $l=52\,\mu\text{m}$ it turns out
$A_1/A_0=108.16$.  The solid line in Fig.~\ref{Fig2} depicts
Eq.~(\ref{E10}) after replacing in it all these constants, together with
the experimental data of Pilling {\it et al}. The agreement is quite
impressive, particularly because it was attained with no adjustable
parameter.

In summary, we obtain close agreement with experiment from interpreting
DB as the interpenetration of the two sufaces at the grain level. The
process is driven by the strong tensile stress field induced in the
plane of the interface by the plastic deformation in the normal
direction. At each point of contact the grain boundaries of one
surface yield to host grains of the other surface, and reciprocally,
releasing this way the internally generated tensile stresses. Voids
close by the gradual increment of the interpenetrated contacting
areas. Hence, bonding is not a matter of contacting and atomic
interdiffusion, but of grain exchange.


\begin{thebibliography}{99}

\bibitem{Hefti} L. D. Hefti, {\it J. Mat. Eng. Perform. \bf 17,}
178 (2008).

\bibitem{Nieh} T. G. Nieh, J. Wadsworth and O. D. Sherby, {\bf
Superplasticity in metals and ceramics} (Cambrige, UK 1997).

\bibitem{Pilling1} J. Pilling, {\it Mater. Sci. Eng. \bf 100,}
137 (1988).

\bibitem{Ridley} N. Ridley, M. T. Salehi and J. Pilling, {\it Mater.
Sci. Technol. \bf 8,} 791 (1992).

\bibitem{Lagos1} M. Lagos, {\it Phys. Rev. Lett. \bf 85,} 2332 (2000).

\bibitem{Lagos2} M. Lagos, {\it Phys. Rev. B \bf 71,} 224117 (2005).

\bibitem{LagosRetamal} M. Lagos and C. Retamal, {\it Phys. Scr.
\bf 81,} 055601 (2010).

\bibitem{LagosRetamal1} M. Lagos and C. Retamal, submitted to {\it
Phys. Scr}.

\bibitem{Qi} Y. Qi and P. E. Krajewski, {\it Acta Mater. \bf 55,} 1555
(2007).

\bibitem{Fukutomi} H. Fukutomi, T. Yamamoto, K. Nonomura and K.
Takada, {\it Interface Sci. \bf 7,} 141 (1999).

\bibitem{Bellon} P. Bellon and R. S. Averback, {\it Phys. Rev. Lett.
\bf 74,} 1819 (1995).

\bibitem{Pilling2} J. Pilling, Z. C. Wang and N. Ridley, in {\it
Superplasticity and Superplastic Forming 1998}, p. 297, ed. by A. K.
Ghosh and T. R. Bieler (The Minerals, Metals and Materials Society,
Warrendale, PA, 1998).

\end{thebibliography}
\end{document}